\documentclass[a4paper,fleqn]{cas-sc}

\usepackage[english]{babel}
\usepackage[capitalize,noabbrev]{cleveref}
\usepackage{csquotes}
\usepackage[numbers]{natbib}
\begin{document}

\shorttitle{RLHF for Human-Robot Collaboration}
\shortauthors{A. Coroiu et~al.}

\title[mode=title]{A Scoping Review and Experimental Study on Reinforcement Learning from Human Feedback for Human-Robot Collaboration}

\author[1]{Alexandra Coroiu}\cormark[1]\ead{alexandra.coroiu@dlr.de}
\author[1]{Andrea Vogt}
\author[2]{Viktor Werbilo}
\author[1]{Andreas Poppele}
\author[2]{Johann Christensen}
\author[2]{Sven Hallerbach}

\affiliation[1]{organization={DLR Institute for AI Safety and Security},
             addressline={Wilhelm-Runge-Straße 10},
             postcode={89081},city = {Ulm}, country={Germany}}
\affiliation[2]{organization={DLR Institute for AI Safety and Security},
             addressline={Rathausallee 12},
             postcode={53757}, city = {Sankt Augustin}, country={Germany}}

\cortext[1]{Corresponding author}

\orcidauthor{0009-0005-6643-7306}{Alexandra Coroiu}
\orcidauthor{0009-0008-9203-2312}{Viktor Werbilo}
\orcidauthor{0000-0001-9871-122X}{Johann Christensen}
\orcidauthor{0009-0007-9163-0567}{Sven Hallerbach}

\begin{abstract}
Human-Robot Collaboration (HRC) can facilitate mass customisation in Industry 4.0, with Reinforcement Learning from Human Feedback (RLHF) representing a promising approach for developing safe AI-based robots.
Practical challenges remain regarding safety during AI development, human feedback quality, and bidirectional human-robot adaptation.
We conducted a scoping review of RLHF in HRC systems, mapping methods that address these challenges. Following PRISMA guidelines, we screened 199 records and included 20 peer-reviewed publications (2020-2025) spanning multiple HRC domains.
To our knowledge, this is the first review focused on the bidirectional, closed-loop design of RLHF.
Our review found multiple feedback modalities enabling data collection in various feedback formats.
Collected data can be integrated at different stages of AI training, resulting in a multi-step development process.
Pilot experiments are commonly used to evaluate HRC systems based on both human and robot metrics.
To empirically test a key gap identified in the review, we conducted a between-subjects VR experiment comparing system- and user-initiated feedback on robot proxemic behaviour for safe navigation. 
Using Bayesian models, we analysed the relation between the collected feedback and safety metrics: psychological safety (post-experiment questionnaire) and physical safety (inverse time-to-collision).
Results show that user-initiated feedback captures perceived safety better than system-initiated feedback, indicating that feedback timing directly affects feedback quality.
Our review and experiment findings show that RLHF relies on appropriate feedback methods to ensure AI safety in HRC, and future RLHF research should prioritise realistic HRC experiments evaluating the effects of feedback collection methods on relevant human and robot metrics.
\end{abstract}

\begin{keywords}
Human-Robot Collaboration \sep Human-in-the-Loop \sep Reinforcement Learning from Human Feedback \sep AI Safety \sep Human-Robot Proxemics
\end{keywords}

\maketitle

\section{Introduction}
%
\subsection{Background}
%
Industry 4.0 has seen a surge of increasingly autonomous robots working alongside humans across various sectors, particularly in manufacturing and logistics applications.
Industrial environments are transitioning from traditional isolated workcells towards Human-Robot Collaboration (HRC), where humans and robots work in shared spaces.
These HRC systems combine the efficiency of automation with human flexibility, enabling robots to operate in dynamic environments~\cite{Liu.2024, Zacharaki.2020}.
Such collaborative robots can be deployed for a wide range of industrial tasks: mobile robots enable navigation and transport on the work floor, and robotic arms enable variable pick-and-place operations for manipulation and assembly~\cite{Lu.2022, Bi.2021}.
By rapidly automating a wide range of repetitive tasks, HRC systems facilitate working with smaller, customised product batches, and allow humans to focus on higher-level decision-making.
This combination of efficiency and adaptability can help address the growing need for mass customization~\cite{Giallanza.2024, Bi.2021}, offering a competitive advantage over traditional automation.

\subsection{Motivation} 
%
However, without barriers to separate humans from robots, safety risks pose a significant challenge for the implementation of HRC systems in industrial settings~\cite{Giallanza.2024}.
Specifically, collisions and non-functional contacts between humans and robots represent sources of injury in the shared workspace~\cite{Gualtieri.2021, Vicentini.2020}.
This risk is particularly critical for industrial robots capable of generating high forces.
Safety strategies focus on preventing hazardous contact (pre-collision strategies) and limiting the impact of the contact (post-collision strategies)~\cite{Hjorth.2022, Zacharaki.2020}.
A series of ISO standards apply to HRC and provide guidelines for the development of safe collaborative robots.
ISO 10218-2~\cite{ISO10218-2} prescribes safety requirements based on the interaction characteristics between the human and the robot.
Specifically, collaborative robots require inherent safety features, such as power and force limiting.
The main practical challenge in developing safety methods for HRC concerns ensuring ISO compliance while preserving high system performance~\cite{Vicentini.2020}.
Therefore, balancing safety and performance becomes an optimisation problem.

Besides immediate hazards like collisions, human factors represent an important latent safety concern in HRC.
Human-centred design of collaborative robots should account for physical and psychological human capabilities, not only to reduce hazard risks and strain for workers~\cite{Cardoso.2021}, but also to ensure optimal system performance~\cite{GonzalezSantocildes.2024, Neumann.2021, Vicentini.2020}.
The most notable psychological factors for HRC are trust, cognitive workload, stress, and perceived safety~\cite{Lorenzini.2023, Hopko.2022}.
Previous research~\cite{Giallanza.2024} identifies cognitive factors as the \enquote{most promising research challenge} for developing safe collaborative robots. The mental processes involved in HRC (i.e., perception, memory, reasoning, motor response)~\cite{Cardoso.2021, Benos.2020} need to be carefully considered in order to neither overload nor underload workers when collaborating with robots on shared tasks~\cite{Hopko.2022}.
These human factors are particularly relevant for ensuring safety and performance over prolonged periods of interaction with collaborative robots.

Currently, there is an ongoing effort to integrate human factors in the development process of safe collaborative robots.
The main goals are to design appropriate interaction modalities for HRC~\cite{Liu.2024, Lu.2022, Hjorth.2022}, and to develop adaptable robots that can accommodate the needs and limitations of workers~\cite{GonzalezSantocildes.2024, Liu.2024, Lu.2022, Cardoso.2021, Gualtieri.2021}.
Existing industrial collaborative robots typically rely only on predefined safety rules, lacking the ability to process real-time human data and dynamically adapt.
Processing high volumes of human data remains an open challenge, due to the difficulty in interpreting complex multimodal data and extracting relevant information about human states.
Solving this challenge requires new technical solutions capable of combining a variety of sensor data and computing safe robot actions accordingly.
Therefore, the development of an adaptive collaborative robot solution relies on measurement methods developed for real-time monitoring of human factors in the real workspace~\cite{Lorenzini.2023, Hopko.2022, Lu.2022}, and advanced artificial intelligence (AI) methods that can improve robot understanding of its environment, including human agents~\cite{Zacharaki.2020, Benos.2020}.
Additionally, these monitoring and AI methods must comply with existing legal and ethical frameworks, such as the the European AI Act~\cite{AIAct.2024} and General Data Protection Regulation~\cite{GDPR.2016}.

\subsection{Problem Statement }
%
The field of HRC requires new AI engineering methods to develop safe collaborative robots that can adapt to complex human environments.
Human-in-the-loop (HITL) approaches constitute one potential solution that helps ensure safety-by-design by directly integrating human input in the AI development process~\cite{Zacharaki.2020}.
A promising AI method is reinforcement learning, which allows robots to learn from continuous interaction with their environment, including human coworkers~\cite{GonzalezSantocildes.2024, Giallanza.2024}.
The use of HITL methods for reinforcement learning relies on diverse interaction modalities (e.g., vision, motion, language) to collect human data that is used as feedback for RL algorithms~\cite{PuertaBeldarrain.2025, Chen.2025}.
These methods can support two main robot learning techniques: imitation learning and reinforcement learning from human feedback (RLHF)~\cite{Chen.2025}.
Imitation learning allows a robot to quickly learn a new task from human demonstrations (e.g., using recorded video samples or hand-guiding data), facilitating human-to-robot skill transfer.
RLHF helps to iteratively refine a learnt task based on human feedback (e.g., through pairwise preferences or ratings), without the need to design a rule-based reward model.
RLHF can be especially useful in complex environments where robots need to adapt to workers' subjective evaluations of safety.
Therefore, RLHF can contribute to increased HRC safety by integrating human factors into AI training.

While RLHF represents a promising approach for developing adaptive collaborative robots, there are still practical challenges that need to be overcome before it can be confidently applied in real-world HRC systems.
Firstly, an important aspect of developing HRC with RLHF methods is ensuring human safety during the development of the AI model.
For this purpose, simulations have been identified as a safe and flexible tool to collect human data for training and testing the AI model before the AI-based robot is deployed in physical environments~\cite{GonzalezSantocildes.2024}.
However, while simulations are useful in the early stages of development, further testing is necessary for validating the AI-based robot within real-world constraints.
Secondly, all RLHF approaches rely on high-quality human data, which is limited by inherent uncertainty (e.g., unpredictable behaviours and human errors), as well as the difficulty of collecting enough data.
To address this, human digital twins can be used in simulation to model human behaviour for training the AI-based robot~\cite{GonzalezSantocildes.2024, Giallanza.2024}.
Nevertheless, collecting real human data remains essential for customising and validating the safety and performance of these methods in HRC applications.
Lastly, RLHF approaches are characterised by the bidirectional adaptation between the human and the robot.
As the robot adapts to human feedback, the human responds to the robot.
This creates a non-stationary learning environment, where iterative feedback loops lead to reciprocal learning~\cite{Nixdorf.2026}.
Therefore, in order to ensure optimal HRC system safety and performance, a closed-loop design is needed to account for both directions of the HITL approach: human-to-robot and robot-to-human~\cite{Hopko.2022}.
Our work contributes towards closed-loop RLHF design with a scoping literature review on state-of-the-art methods, restricted to bidirectional studies, and a VR experiment comparing the effects of two feedback collection conditions on human factors and feedback quality.

\section{Literature Review}
\subsection{Research Questions}
\begin{figure}[pos=htb]
    \centering
    \includegraphics[width=0.7\textwidth]{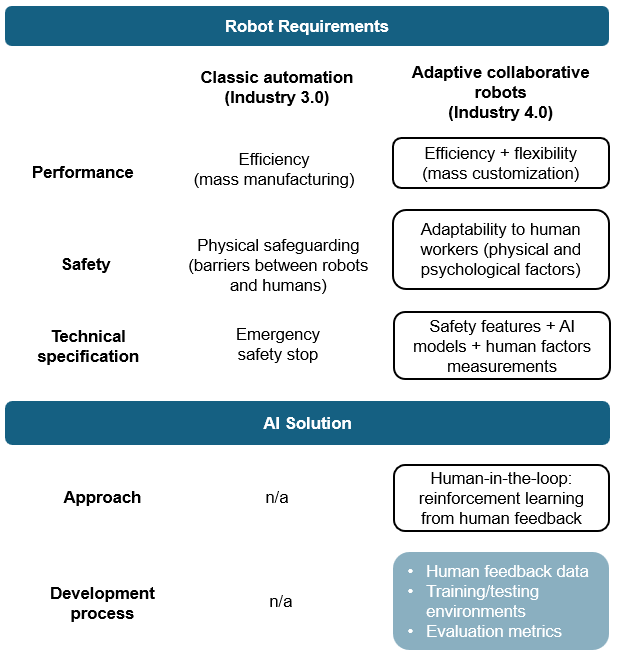}
    \caption{Requirements for adaptive collaborative robots compared to classic automation, and RLHF as a potential solution}
    \label{scope}
\end{figure}

As industry shifts towards higher levels of autonomy and mass customisation, robots face increasing performance demands both in terms of efficiency and flexibility.
HRC systems can help meet these performance needs, but also introduce new safety concerns in the workspace.
While classic automation uses rigid constraints to ensure safety, collaborative robots should dynamically adapt to human workers to balance the safety-performance trade-off (see \cref{scope}).
HITL approaches represent a promising solution for developing such adaptive collaborative robots.
In particular, RLHF methods provide a promising paradigm to directly integrate human data into robot learning.
However, the current development of RLHF methods for collaborative robots is still limited by methodological challenges regarding safety during AI development, the complexity of human feedback data, and the effects of human-robot co-adaptation.
Addressing these challenges is crucial for developing RLHF applications that can be safely deployed in real-world HRC systems.
Therefore, this scoping literature review aims to investigate the state-of-the-art methods used for RLHF development in HRC systems, focusing on solutions that account for the bidirectional adaptation between the human and AI-based robot.
Accordingly, we structured this review around the following research questions (RQs):
\begin{itemize}
    \item \textbf{RQ1:} What HRC systems employ RLHF models for robot task learning? \label{RQ1}
    \item \textbf{RQ2:} What feedback methods are used to collect human data for RLHF models in HRC systems? \label{RQ2}
    \item \textbf{RQ3:} What AI development methods are used for training RLHF models in HRC systems? \label{RQ3}
    \item \textbf{RQ4:} What metrics are used for evaluating the effects between the human and the RLHF-trained robot in HRC systems? \label{RQ4}
\end{itemize}

\subsection{Methods}
\begin{figure}[pos=htb]
    \centering
    \includegraphics[width=0.75\textwidth]{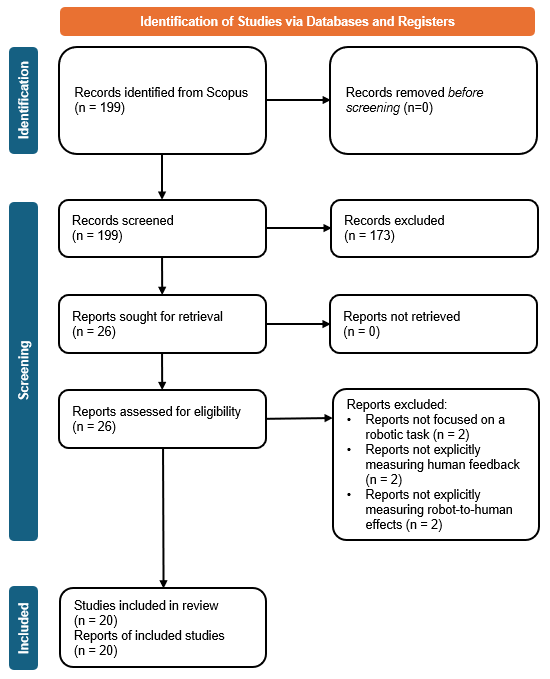}
    \caption{PRISMA flow diagram~\cite{Prisma.2021} of the Scopus search (2020-2025), 199 records screened, 20 included}
    \label{prisma}
\end{figure}

To conduct this scoping literature review we searched for relevant publications using the following search string: \texttt{(robot* OR cobot*) AND ("human in the loop" OR "human feedback" OR "user feedback" OR "human reward" OR "user reward" OR "human preference*" OR "user preference*") AND ("reinforcement learning" OR rl OR "preference learning" OR rlhf)}.
We queried the Scopus database for journal articles and conference papers published between 2020 and 2025, reflecting the emergence of RLHF methods in robotics.
The search resulted in 199 records.
We screened the title and abstract of these records to select reports that match our research scope (see~\cref{prisma}).
Each report was screened by two reviewers independently.
We used the following relevance criteria:
\begin{itemize}
    \item RLHF is used for learning a robot specific task (e.g., navigation, manipulation); excluding general AI tasks (e.g., language processing, visual processing).
    \item Humans directly participate in robot learning, beyond initial AI algorithm development (e.g., not limited to designing the reward model).
    \item The AI-based robot integrates human data as feedback for training the RLHF algorithm.
    \item Experiments involve real human participants; excluding the use of purely synthetic/simulated human data.
    \item Metrics are evaluated for both directions of the HITL: human-to-robot effects and the robot-to-human effects.
\end{itemize}
Consistent with our research questions, we recorded the following data items from each included report: application domain, specific robot task, human data measurement, feedback format, experiment environment, AI development methods, human metrics (cognitive factors and feedback behaviour), AI-based robot metrics.

\subsection{Results}
\subsubsection{RLHF Applications}
%
The use of HITL approaches spans across a wide range of robotic applications.
RLHF is used for training robots to fulfil roles in industry~\cite{Avaei.2023, Tahara.2025, Yousefi.2025}, assistance and care~\cite{Hiranaka.2023, KumarShastha.2020, MarotoGomez.2024, McQuillin.2022, Schneider.2021, Wang.2022, Wang.2025} or emergency and disaster situations~\cite{Huang.2021}.
Depending on the type of robot, different tasks can be trained with RLHF.
Mobile robots can be fine-tuned for navigating in busy environments where robot and human paths can easily cross~\cite{Kazantzidis.2022, Marta.2022, Marta.2023, McQuillin.2022, Raza.2019, Tahara.2025, Wang.2022}.
Robotic arms can be trained to perform complex tasks that require fine movements in proximity to humans.
Human feedback is then used to fine-tune the RLHF policy for robotic arm navigation in a 3D space~\cite{Avaei.2023, Hindemith.2023, Kim.2020, Yousefi.2025, Yu.2023, Zhang.2023} or for the control of the robotic grippers~\cite{Hiranaka.2023, KumarShastha.2020, Wang.2025}.
These RLHF use cases can highly benefit from close collaboration between humans and robots to enhance capabilities for flexible tasks in uncertain environments.

\subsubsection{Human Feedback Methods}
\begin{figure}[pos=htb]
    \centering
    \includegraphics[width=0.95\linewidth]{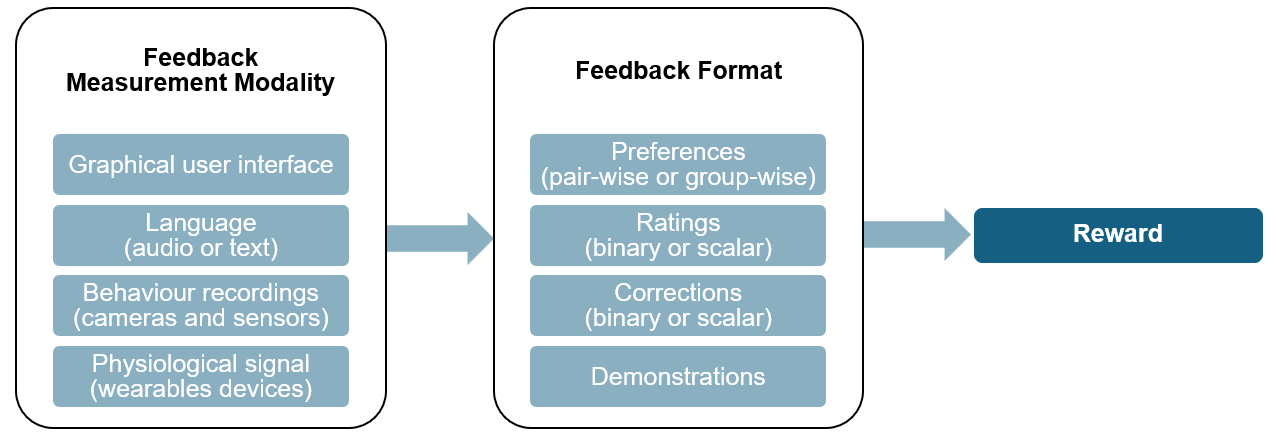}
    \caption{Feedback methods in RLHF for robotic applications: measurement modalities and formats}
    \label{feedback}
\end{figure}

%
There is a wide variety of proposed solutions for collecting human feedback to be used in RLHF algorithms (see \cref{feedback}).
These solutions encompass both direct and indirect feedback measurements, each requiring different modalities and different levels of data pre-processing.
One of the most common feedback methods involves using a graphical user interface to collect subjective evaluations of the interaction with the robot~\cite{Hindemith.2023, Hiranaka.2023, Marta.2022, Yu.2023}.
Language-based feedback represents another, more complex form of subjective measurement~\cite{Wang.2025}.
This method requires an intermediate natural language processing step, e.g., to evaluate the message or emotional valence in verbal or spoken text.
Indirect, objective feedback measurements also require more advanced data processing to evaluate complex input from user behaviour~\cite{Yousefi.2025} (captured through cameras and sensors) or physiological signals like brain-computer interfaces~\cite{Kim.2020}.
Interpreting these signals relies on statistical or AI-based analysis to extract meaningful feedback information.
Lastly, multi-modal setups can be used to collect robust feedback from combinations of various measurements~\cite{MarotoGomez.2024, McQuillin.2022}.

The types of feedback collected through the different measurement methods can vary significantly (see \cref{feedback}).
First, simple feedback formats consist of subjective preferences~\cite{Byk.2022, Kazantzidis.2022, Marta.2023, Schneider.2021, Wang.2022, Zhang.2023, Hindemith.2023} and binary or scalar ratings~\cite{Hiranaka.2023, MarotoGomez.2024, Marta.2022, McQuillin.2022, Yu.2023, Zhang.2023, Hindemith.2023} which can be more easily collected through GUIs.
However, a key consideration in presenting preferences to users is the need to pre-select relevant pairs or groups of items for comparison~\cite{Zhang.2023, Byk.2022}.
Second, corrections can be used to provide negative feedback for undesirable robot behaviours either as simple binary input or as a scalar value as a measure of intensity~\cite{KumarShastha.2020, Tahara.2025, Wang.2025, Huang.2021}.
Lastly, feedback from demonstrations~\cite{Avaei.2023} represents a more complex format that requires more advanced user behaviour input (e.g., from cameras, force-feedback sensors) and data processing to compare the desired actions against the robot motion and compute feedback values from this difference.
As with the different feedback measurements, feedback formats can also be combined to acquire more robust feedback data~\cite{Yousefi.2025, Raza.2019, Kim.2020}. 

Regardless of feedback measurement or format, collected data ultimately needs to be converted into a reward score that can be used in the RL algorithm.
The key challenge with human feedback is the inherent uncertainty within human data.
While some noise is expected, the design and implementation of feedback methods can influence the noise levels in multiple ways~\cite{Marta.2022, Yousefi.2025, Zhang.2023}.
For example, \cite{Yu.2023} compares the noise levels between binary and scalar ratings.
Similarly, when it comes to preferences, the specific options presented can also highly impact human feedback~\cite{Byk.2022}.
Furthermore, complex methods that require extensive data pre-processing make it harder to isolate the specific factors that shape the final reward score distribution. Another challenge with human feedback is collecting sufficiently large and heterogeneous datasets to create representative distributions that are not biased towards a certain type of feedback response~\cite{Avaei.2023, Marta.2023}.

\subsubsection{AI Development Methods}
\begin{figure}[pos=htb]
    \centering
    \includegraphics[width=0.7\textwidth]{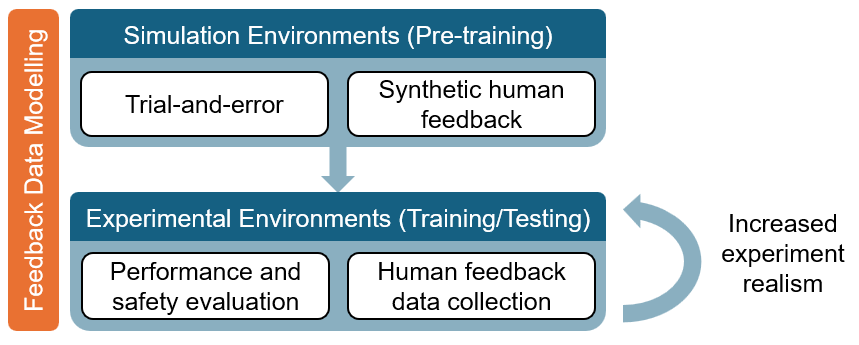}
    \caption{RLHF development methodology for robotic applications with iterative training stages}
    \label{methods}
\end{figure}

%
Developing RLHF algorithms for HRC generally requires a multi-step process to achieve the desired AI-based robot learning performance (see \cref{methods}).
Using pre-training for the RL task before collecting human feedback~\cite{Kim.2020, Schneider.2021, Wang.2022, Wang.2025, Yousefi.2025} ensures a functioning initial policy that can then be fine-tuned and customised according to user input through subsequent learning sessions.
RL algorithms can be pre-trained using trial-and-error approaches in simulation, allowing the robot to learn independently, or by incorporating synthetic human feedback to introduce human evaluation early in the learning process~\cite{Huang.2021, Marta.2022, Marta.2023, Wang.2025}.
These methods enable training on a large quantity of trials and data that would otherwise be difficult, if not impossible, to collect through real interaction with the environment and users.
For modelling both the synthetic and the real human feedback, Bayesian methods can be employed to create distributions of the human data and account for feedback uncertainty~\cite{Marta.2022, Yousefi.2025, Zhang.2023}.

Experiments for collecting human feedback can be set up in various environments that provide different levels of realism.
On the one hand, 2D or 3D computer simulations offer an easy-to-use framework for displaying AI-based robot behaviour and collecting human feedback~\cite{Huang.2021, Marta.2022, Marta.2023, Raza.2019, Yu.2023, Kazantzidis.2022}.
On the other hand, realistic (mock-up) environments require more space and functional physical assets~\cite{Kazantzidis.2022, McQuillin.2022, Wang.2022, Wang.2025, Yu.2023}, but offer greater realism and the opportunity to collect more representative feedback data.
Additionally, an important consideration for meaningful experiments in realistic environments is the participant count and background.
Pilot studies are carried out with a relatively small number of participants ($<25$)~\cite{Huang.2021, Kim.2020, KumarShastha.2020, MarotoGomez.2024, McQuillin.2022, Tahara.2025, Wang.2022, Wang.2025, Yousefi.2025, Zhang.2023}, while fewer large-scale studies exist~\cite{Marta.2022, Marta.2023, Raza.2019, Schneider.2021, Yu.2023}.
Furthermore, although these large-scale studies include more participants, they are not representative of the intended target groups.

\subsubsection{Evaluation Metrics}
\begin{figure}[pos=htb]
    \centering
    \includegraphics[width=0.75\textwidth]{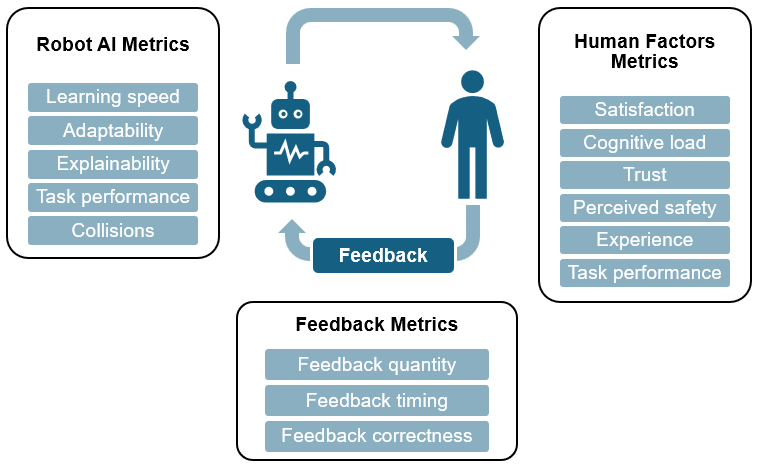}
    \caption{HRC safety and performance for the closed-loop RLHF design: AI-based robot, human and feedback metrics}
    \label{factors}
\end{figure}

%
The human-to-robot direction within HTL concerns the effects on the AI-based robot and collaboration outcomes (see \cref{factors}).
Within the context of RLHF for HRC applications, AI-based robot performance can be evaluated on two levels.
Firstly, AI learning properties determine the suitability of the RL model for the intended task and target users.
AI learning can be quantified using quality metrics such as learning efficiency~\cite{Hiranaka.2023, Huang.2021, Kazantzidis.2022, Marta.2023, Marta.2022, Wang.2022, Wang.2025}, often assessed alongside adaptability~\cite{Schneider.2021, McQuillin.2022, Avaei.2023, Kim.2020, KumarShastha.2020}, or explainability (especially in terms of predictability)~\cite{Hindemith.2023}.
These metrics represent AI characteristics necessary for enabling the robot to perform highly flexible tasks.
Secondly, the robot's resulting behaviour directly impacts overall HRC system outcomes.
While most studies focus only on robot task performance metrics (e.g., efficiency in task execution)~\cite{Hindemith.2023, KumarShastha.2020, Marta.2022, Tahara.2025, Yousefi.2025, Yu.2023, Zhang.2023}, others specifically focus on robot actions that impact system safety (e.g., collisions in navigation)~\cite{Avaei.2023, Wang.2022, Hiranaka.2023, Huang.2021}.

The robot-to-human direction within HTL concerns the effects on human factors and collected feedback (see \cref{factors}).
Within the context of RLHF for HRC applications, two main aspects must be considered for evaluating human factors.
Firstly, psychological factors play a central role in evaluating the human-robot dynamic.
The most frequently studied factor is satisfaction~\cite{Byk.2022, KumarShastha.2020, McQuillin.2022, Wang.2022, Yousefi.2025} (typically evaluated using the System Usability Scale questionnaire~\cite{Brooke.1996}) often assessed alongside cognitive load~\cite{Avaei.2023, Hindemith.2023, Raza.2019, Tahara.2025, Zhang.2023} (typically evaluated using the standard NASA TLX questionnaire~\cite{NASA-TLX}).
Trust~\cite{Schneider.2021} and perceived safety~\cite{Wang.2025} are only measured in one study each.
Additionally, a small number of studies account for user experience and learning effects~\cite{Hiranaka.2023, Yu.2023, Schneider.2021, Tahara.2025}.
Secondly, behavioural aspects encompass both the primary task performance~\cite{MarotoGomez.2024, Yu.2023} and the provision of human feedback, which can be considered a secondary task in the context of HRC.
The most studied human feedback metric is the total number of collected data points~\cite{Avaei.2023, Hiranaka.2023, Huang.2021, Kazantzidis.2022, Kim.2020, Marta.2022, Tahara.2025, Yousefi.2025}, which is highly relevant for developing a robust reward function in the RL process.
Additionally, it can also be relevant to evaluate feedback timing~\cite{Zhang.2023} or feedback correctness~\cite{Kim.2020, Raza.2019}. 

\subsection{Discussion}
\subsubsection{Main Findings}
%
%
For RQ1, the results reveal that a wide range of HRC applications employ RLHF for robot training: from navigation AI for disaster robots to manipulation AI tasks for assistive robot arms.
The common ground among all these applications is the need to operate despite high uncertainty, either because the area is unknown (e.g., in disaster sites), users are diverse (e.g., in assistive care), or tasks change frequently (e.g., mass customisation in industry).
These applications often also require very close collaboration with humans: assistive robots work in close physical contact with humans, while emergency or industrial robots need to complement human efforts.
Therefore, any HRC system that requires flexibility can benefit from RLHF approaches to optimise the AI-based robot using human input.

For RQ2, the results reveal a wide range of state-of-the-art technical solutions for integrating human feedback in RL algorithms.
The research appears to be in an exploratory phase where various interaction tools (classic GUIs, NLP, cameras and sensors, or even BCIs) are used to collect feedback in various formats (e.g., ratings, preferences, corrections, or demonstrations).
There are no universally preferred methods, as the feedback requirements can vary by application use case or AI development stage.
A common issue arising across feedback methods is data uncertainty, stemming from noise and bias in human responses.
Feedback strategies are often focused on maximising feedback input counts, maximising information gain from input data points, or minimising uncertainty. 

For RQ3, the results show that human feedback alone cannot be directly used to produce desirable RL model performance.
The AI development methodology often requires pre-training in simulation and synthetic human data to train the robot initially.
Additionally, when training with real human feedback data, AI methods can benefit from Bayesian modelling to represent noisy feedback distributions.
Furthermore, the collection of real human feedback currently depends on low-fidelity prototypes of the intended HRC system (computer simulations or mock-up environments), which are often used to run pilot studies.
These findings suggest that the development of RLHF methods requires iterative steps to train and test the AI, from low-level simulations in the first steps to realistic environments with real users in the later steps. 

For RQ4, several human and AI-based robot metrics have been identified for HRC systems.
For the AI-based robot, metrics which reflect flexibility are most relevant: how well and how fast the AI model can learn to adapt to new users, tasks or environments.
These AI metrics shape the overall robot behaviour, subsequently affecting HRC system outcomes.
When it comes to human behaviour, the research is predominantly focused on the quantity of provided feedback, which is highly relevant within the context of RLHF.
In terms of psychological factors, existing research is mainly focused on satisfaction, while the only safety-relevant metric commonly evaluated is cognitive load.
This highlights the early exploratory nature of RLHF research in HRC, which currently focuses more on AI development rather than overall system safety and performance. 
\subsubsection{Limitations}
Two main limitations can be noted.
First, this scoping literature review sits at the intersection of two related, but distinct research fields: Human-Robot Interaction and AI development with HITL methods.
Constructing a search string that covers relevant RLHF methods applicable to HRC systems proved to be difficult.
This is an emerging field, so standard terminology is not well established yet.
HRC literature focuses on human factors and robotics using vocabulary specific to the psychology and engineering domains, whereas HITL-RLHF literature focuses on AI development using technical vocabulary specific to AI model optimisation and evaluation.
Our search string captures keywords specific to our research scope, but may have omitted relevant reports that use other terminology.
Furthermore, we restricted our search to the Scopus database due to its multidisciplinary coverage of both HRC and AI literature, which is well suited for this emerging field, but publications indexed only in other databases may have been missed.
Second, we restricted our selection criteria to studies that included both directions of the HITL approach, assuming that a closed-loop design requires specific methods to capture the bidirectional adaptation between human and AI-based robots.
Nevertheless, unidirectional studies might also contain applicable feedback and AI development methods, but these methods should be validated for applicability within bidirectional studies.

\subsubsection{Future Research}
%
RLHF research for HRC can benefit from focusing solutions on target applications.
When it comes to feedback formats and collection methods, tailored solutions should be developed for specific use cases~\cite{Lorenzini.2023, Lu.2022}.
For example, an industrial environment with specific HRC tasks might impose various constraints on the RLHF methodology, such as environment sound and light conditions, monitoring capabilities, available interaction modalities, time and cognitive limitations, procedural standards, resource availability.
These constraints limit the viability of different feedback methods and should be considered early in the design and development of RLHF strategies.
Therefore, future research should attempt to integrate and test in realistic set-ups with target user groups. 

Another aspect of bringing RLHF research closer to real-world applications of HRC systems is focusing on system safety metrics.
Currently, there is limited attention given to human factors that impact HRC safety: trust, perceived safety, and stress~\cite{Lorenzini.2023, Hopko.2022}.
These human factors should be evaluated in realistic environments, under different experimental conditions, and with varied feedback strategies.
Particularly, cognitive load can be further explored, as the secondary feedback task can affect both safety and the overall outcome of the primary collaborative task.
All these psychological factors are relevant not only for avoiding hazards in a shared human-robot workspace, but also for ensuring optimal HRC system performance.
Furthermore, safety metrics should be evaluated in realistic experiments that best reflect the target HRC application.  

Finally, the closed-loop design~\cite{Hopko.2022} of RLHF methods would benefit from evaluating the impact of human-robot collaboration on the collected feedback.
Human feedback behaviour can be influenced by both psychological factors and real-world system properties (e.g., environmental conditions, task requirements, available feedback methods), resulting in varying levels of noise or bias.
Similarly, the AI-based robot behavioural policy, defined by prior training phases (e.g., pre-training on synthetic human data), ultimately shapes the interaction with the human and the characteristics of the provided feedback.
These factors should be accounted for when modelling feedback as rewards for RLHF in order to reduce uncertainty in the collected human data.
A well-structured feedback strategy should capture multiple dimensions of the collaboration, including psychological and physical safety. 

\section{Experiment}
\subsection{Research Questions}

RLHF applications in HRC systems rely on human data for training high-performing and safe collaborative robots.
Collected feedback is characterised by both volume and quality of data.
First, a high volume of feedback input is beneficial for model training, but increases effort for the human collaborator.
Therefore, RLHF applications aim to have the robot learn as fast as possible with as little feedback as possible~\cite{Avaei.2023, Hiranaka.2023, Zhang.2023, Kazantzidis.2022, Marta.2022, Huang.2021, Kim.2020}.
Feedback volume can be measured not only in terms of counts, but also as time spent per feedback query~\cite{Kazantzidis.2022}.
Additionally, human effort can be directly quantified as cognitive load~\cite{Avaei.2023, Huang.2021, Raza.2019, Zhang.2023}.
Second, feedback quality can be broadly defined in terms of information value for the RLHF reward model.
Two key considerations that impact feedback quality are correctness~\cite{Kim.2020, Raza.2019} and timing~\cite{Zhang.2023}.
Information value per feedback point can be maximised by designing relevant feedback queries~\cite{Huang.2021, Marta.2022}.

The ideal feedback strategy for RLHF requires optimising the trade-off between human effort and information value obtained from feedback interactions to ensure practicality for real-world HRC systems.
Binary ratings represent a fast, low-effort feedback strategy, but the lower resolution of binary signal requires more feedback queries to train the RL policy~\cite{Hwang.2024}.
The demand for higher feedback volume could be addressed through system-initiated feedback queries at a predetermined sampling rate.
In this study, we want to explore the difference between system-initiated and user-initiated binary feedback.
To address both directions of a closed-loop design for HRC~\cite{Hopko.2022}, we want to evaluate both the effects of feedback condition on the human and the effects of feedback on RLHF robot learning.
Therefore, we conducted a VR experiment to answer the following research questions: 
\begin{itemize}
\item \textbf{RQ5:} How do system-initiated versus user-initiated binary feedback strategies compare in terms of information value?
\item \textbf{RQ6:} How do system-initiated versus user-initiated binary feedback strategies compare in terms of human effort?
\end{itemize}

\subsection{Methods}
\subsubsection{Application Domain}
The experiment focuses on comparing feedback conditions for RLHF methods applied to robot navigation tasks in HRC systems.
Previous research~\cite{Marta.2022, McQuillin.2022} successfully employed RLHF to integrate human proxemic zones into robot navigation and improve human factors outcomes such as perceived safety.
This is particularly relevant for industrial applications where repeated proxemic violations can decrease psychological safety for workers, which in turn degrades overall HRC performance and safety~\cite{Neumann.2021}.
Therefore, safe robot navigation in HRC systems should consider both physical and psychological safety.
Furthermore, proxemic zones are not static, but dynamically change based on robot navigation speed and interaction context~\cite{Neggers.2022}.
RLHF solutions can more accurately capture these dynamic proxemic zones based on human input, instead of relying on AI navigation models with static proxemic rules, resulting in better safety-performance trade-offs.

\subsubsection{Environment and Task}
\begin{figure}[pos=htb]
    \centering
    \includegraphics[width=0.75\textwidth]{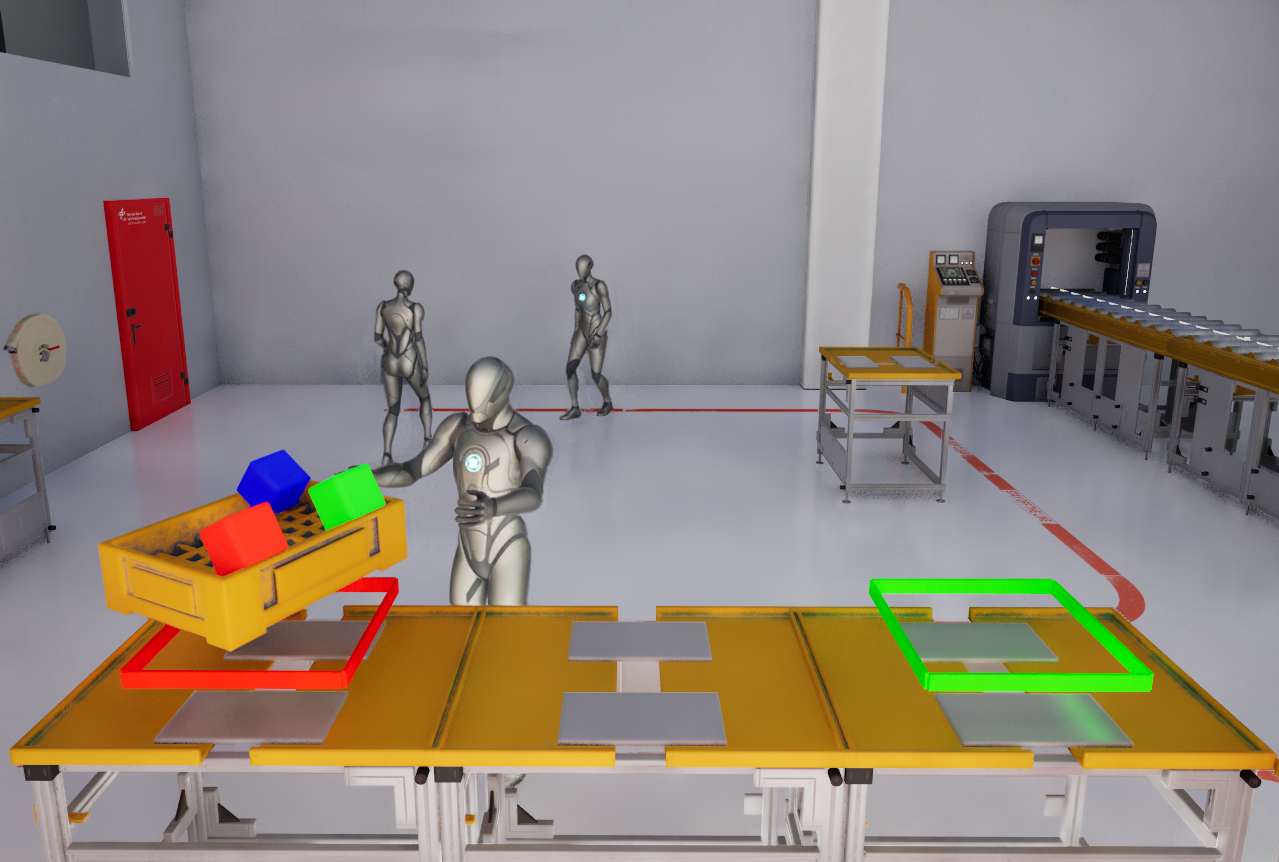}
    \caption{First-person perspective of the VR environment, including the human workstation, co-working robot and background robots}
    \label{VR}
\end{figure}

%
The experiment was conducted in a VR environment developed in Unreal Engine (UE) version 5.4, and displayed on an HTC Vive headset to participants.
The virtual environment simulated an industrial workshop area featuring a sorting station and a conveyor belt (see \cref{VR}).
The environment featured three humanoid robots: one co-working robot and two additional robots operating in the background.
The co-working robot was responsible for delivering unsorted crates to the workstation, where the human sorts the crates by object colour using the HTC controller to interact with the virtual objects.
The co-working robot is then responsible for transporting the sorted crates to the conveyor belt and storing empty crates away.
As the co-working robot navigates the workshop area, its path crosses with the other two robots.
Robot navigation was controlled with the UE navigation AI, and collisions were avoided using safety stops when trespassing the 1m radius around other actors. 

\begin{figure}[pos=htb]
    \centering
    \includegraphics[width=0.75\textwidth]{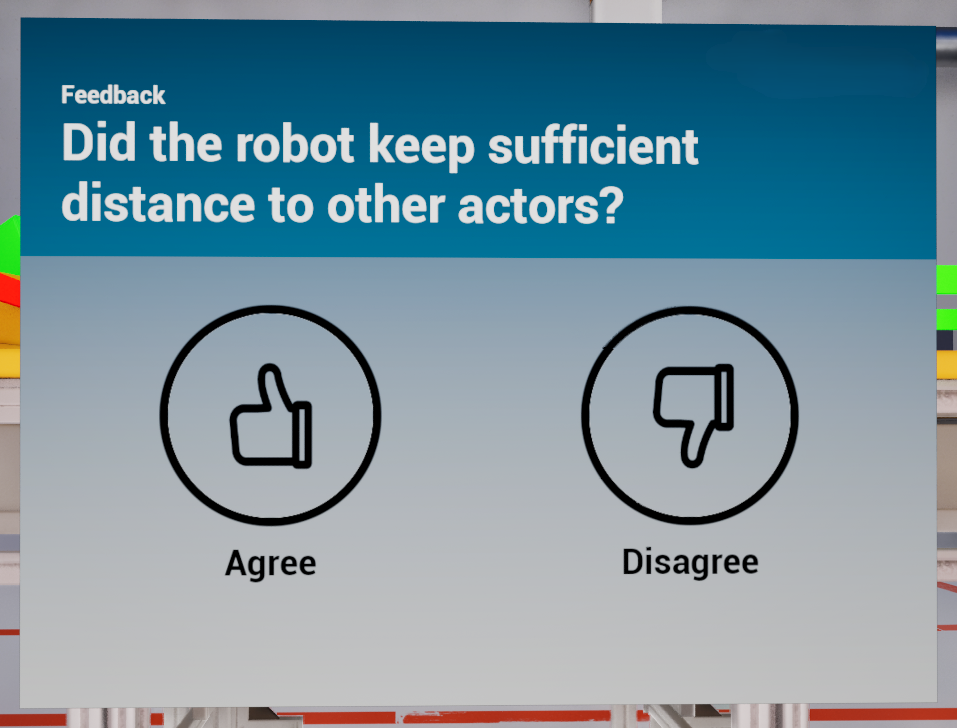}
    \caption{Feedback pop-up prompt displaying the binary feedback buttons}
    \label{prompt}
\end{figure}

The human has to provide feedback on the proxemic behaviour of the co-working robot.
A pop-up feedback interface displays the following question: \enquote{Did the robot keep sufficient distance to other actors?} (see \cref{prompt}).
The human provides binary feedback by choosing between the thumbs-up and thumbs-down buttons.
Providing feedback represents a secondary task to the primary task of sorting crates.
The two feedback conditions are designed to facilitate data collection while maintaining a smooth workflow.
In the user-initiated condition, the participant is free to open the feedback pop-up interface whenever they want with the click of a button on the controller.
In contrast, in the system-initiated condition, the participant is prompted to provide feedback for every third crate transported to the conveyor belt.

\subsubsection{Participants and Procedures}
%
Given the exploratory nature of this experiment, we used a convenience sampling strategy.
Participants were recruited internally at the two institute locations in Germany.
Written informed consent was obtained from all participants, including explanations on data protection and potential VR side effects.
We collected demographic data and VR familiarity (6-item questionnaire) in terms of three dimensions: VR expertise, VR frequency, and VR confidence.
A total of 48 participants completed the experiment, but four were excluded for not providing any feedback during the VR session.
As a result, 44 participants ($Male = 30$, $Female = 13$, $NonBinary = 1$), aged between 25 and 63 years ($M = 33.3$, $SD = 8.8$) were included in the study.
Most participants reported moderate VR expertise ($M = 3.07$, $SD = 1.29$) and confidence ($M = 2.92$, $SD = 0.97$), despite infrequent use ($M = 1.55$, $SD = 1.09$). 
 
The study employed a between-subjects design with two feedback conditions: user-initiated and system-initiated.
Participants were randomly split between the two experimental conditions.
The final distribution of participants between the user-initiated ($N = 23$) and the system-initiated ($N = 21$) feedback conditions was balanced.
To compare the two experimental conditions, we collected data on three safety-relevant psychological factors~\cite{Lorenzini.2023, Hopko.2022}: perceived safety (8 item questionnaire~\cite{Akalin.2019}), cognitive load (NASA TLX~\cite{NASA-TLX}), and sense of control (3 item questionnaire~\cite{Strube.1984}).
The experiment followed a standardised procedure:
\begin{enumerate}
    \item Pre-survey: Participants completed the demographic data and VR familiarity questionnaires.
    \item VR introduction: Participants received an introduction to the VR equipment, environment and tasks.
    \item VR session: Participants performed the task in VR. The session is capped at 15 minutes.
    \item Post-survey: Participants completed perceived safety, cognitive load and sense of control questionnaires.
\end{enumerate}

\subsubsection{Metrics and Analysis}
%
Human effort and feedback information value were quantified based on the survey results and data logs from UE collected during the VR sessions.
First, human effort is defined by two metrics: the average duration (seconds) spent on a feedback prompt in the feedback interface, and normalised cognitive load computed from the post-survey data (in line with~\cite{Avaei.2023, Zhang.2023, Huang.2021}).
Second, feedback information value was defined by two dimensions: psychological and physical safety.
Psychological safety was quantified by the two normalised human factors metrics computed from survey data: perceived safety and sense of control (in line with~\cite{Marta.2022}).
Physical safety was defined by a proxemic threat score quantified as inverse time-to-collision (iTTC = $\text{TTC}^{-1}$) computed from data logs~\cite{Singamaneni.2023}.
We computed threat scores between the co-working robot and the human, and between the co-working robot and the other two robots in the environment.
The maximum threat values were extracted from the five-second window preceding each time the feedback interface was opened and feedback was provided. 

Bayesian models~\cite{Gelman.2013} were fitted to compare the two feedback conditions (see \cref{app:bayes}).
Given the nested data structure, we distinguish participant-level data from feedback-level data in our analysis to account for inter-individual variability and avoid pseudo-replication.
First, to estimate posterior differences in human effort between the feedback conditions, Bayesian models were constructed for each metric: a Gaussian model for cognitive load (participant-level), and a log-gamma model for average feedback duration (feedback-level).
Second, to estimate posterior differences in information value between the feedback conditions, Bayesian models were constructed for each safety dimension: one model for psychological safety and one for physical safety.
Maximum threat to the human and to the other agents were used as predictors in a Bernoulli model with feedback-level data.
The psychological safety metrics were not strongly correlated ($|r| = 0.41$) and could be used as individual predictors in a binomial model with participant-level data. 

\subsection{Results}
\begin{figure}[pos=htb]
    \centering
    \includegraphics[width=0.75\textwidth]{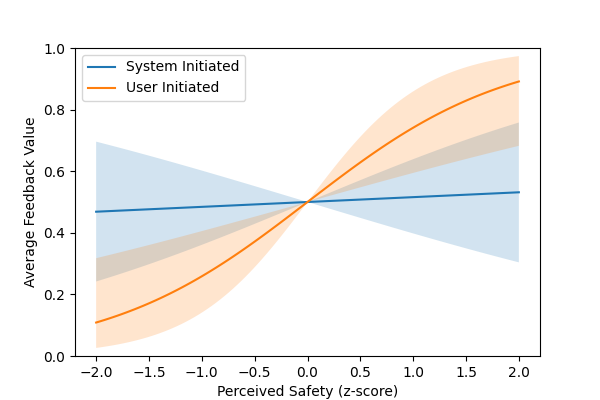}
    \caption{Feedback information value for perceived safety for user- and system-initiated conditions, Bayesian posterior sigmoids (mean and 95\% CrI)}
    \label{fig:safety}
\end{figure}

The system-initiated feedback led to higher feedback volume ($M = 9.22$, $SD = 0.76$) than user-initiated feedback ($M = 5.24$, $SD = 2.4$).
Despite more frequent interaction with the feedback interface in the system-initiated condition, there were no credible differences in cognitive load ($M = -0.15$, 95\% CrI $[-0.78, 0.48]$) or feedback duration ($M = -0.10$, 95\% CrI $[-0.17, 0.39]$) between conditions.
Therefore, the system-initiated condition provides more feedback data without additional human effort.
Feedback valence was generally positive across both conditions, with system-initiated data showing particularly high values ($M = 0.93$, $SD = 0.13$), and user-initiated values being moderately high as well ($M = 0.78$, $SD = 0.28$).
However, there are credible differences in how the two feedback conditions capture psychological safety.
User-initiated feedback reflects perceived safety far better than system-initiated feedback ($M = 1.0$, 95\% CrI $[0.13, 1.9]$) (see \cref{fig:safety}).
This credible interval excludes zero, providing strong evidence of the effect, with model convergence confirmed ($\hat{R} \approx 1.00$).
Additionally, there is also a weak difference in how feedback encodes the maximum proxemic threat posed to the human ($M - 0.16$, 95\% CrI $[-0.01, 0.35]$).
No interaction effects were identified for sense of control or the maximum proxemic threat posed to the other robots.
Full parameter estimates for the psychological and physical safety models are provided in \cref{app:bayes} (\cref{tab:bayes_results_ittc,tab:bayes_results_safety}).

\subsection{Discussion}
The results indicate that user-initiated feedback provides greater information value than system-initiated feedback.
This finding builds on previous research~\cite{McQuillin.2022} demonstrating that real-time user feedback improves robot social behaviour.
Additionally, it extends previous research~\cite{Wang.2022} showing better outcomes with feedback-efficient preference learning.
Despite a lower volume of feedback data points, the user-initiated condition captures more meaningful information about the psychological safety for the human co-worker.
Capturing latent safety constructs is highly valuable for RLHF, because it allows the encoding of not only physical state variables, but also subjective human perception~\cite{Marta.2022, Kazantzidis.2022}.
Additionally, the physical safety data suggests that user-initiated feedback is more sensitive to safety violations.
Therefore, these findings show that user-initiated feedback compensates for the lower data volume by providing more timely and efficient feedback. 

The collected feedback for the system-initiated condition was predominantly positive.
This suggests that the system-initiated condition fails to capture both physical safety violations and psychological safety concerns.
It appears that although human co-workers form a negative impression of perceived safety, this cannot be easily captured without timely feedback action.
When prompted with the feedback interface, participants might respond positively to recent events, even though their overall safety perception is affected by proxemic violations happening outside of the immediate time window.
However, two main limitations to the design of this feedback condition must be noted.
First, the timing of the system-initiated prompt might influence the participants towards a positive response because it overlaps with task completion.
Second, the phrasing of the feedback question might not prompt a strong negative response to proxemic violations.
Because the safety stop prevents collisions, the distance between actors is always technically \enquote{sufficient}, even when it violates proxemic expectations.

The findings from this study can directly inform the design of more realistic future experiments.
Moving towards higher fidelity VR simulation or physical set-up requires new feedback interfaces.
A pop-up prompt similar to what was used in UE can be transferred to a mounted display on a physical workstation.
However, the interaction workflow changes as the human needs to switch attention from the primary physical task to a secondary digital feedback task.
The design of the workflow can impact both human effort and the quality of collected feedback.
Additionally, for system-initiated feedback, a different modality might be necessary for gaining the attention of the human in a busy industrial environment.
More advanced feedback modalities can be employed by having the human give feedback directly to the robot.
This would considerably increase the complexity of the human-robot collaboration loop, but could provide a better effort vs information value trade-off. 

\section{Conclusion}
%
RLHF enables the development of flexible AI-based robot applications across multiple domains where close HRC is required, such as manufacturing and logistics.
Our work provides two contributions to RLHF in HRC: a scoping literature review focused on the bidirectional, closed-loop design of RLHF, and a VR experiment assessing the interplay between human factors, robot behaviour, and feedback quality.
The review shows that RLHF is an emerging research field, currently exploring a wide range of methods for integrating human feedback into robot training during various stages of the AI development process.
Our experimental results indicate that feedback timing is a key aspect in ensuring AI safety, as human data quality depends on immediate feedback action following a perceived safety threat.
Future RLHF research would benefit from conducting studies in high-fidelity realistic settings to achieve the desired safety and performance goals for HRC systems.
Particular attention should be paid to ensuring that feedback methods are viable in real environments, and able to produce high-quality data for RL algorithms without compromising human factors.
Ultimately, RLHF methods have the potential to provide AI-based robots with the needed adaptability to bring HRC systems into highly variable real-world applications like Industry 4.0.

\bibliographystyle{cas-model2-names}
\bibliography{ref_HRC, ref_HItL, ref_RLHF, ref_tools}

\newpage
\appendix
\section{Bayesian Statistics} \label[appendix]{app:bayes}
\counterwithin{table}{section}

Bambi specification for Gaussian model on cognitive load:
\begin{verbatim}
cognitive_load = bmb.Model(
    "NASATLX ~ condition",
    part_df,
    family = "gaussian"
)
\end{verbatim}

Bambi specification for gamma model on feedback duration:
\begin{verbatim}
feedback_duration = bmb.Model(
    "fb_dur ~ condition + (1|pid)",
    df,
    family = "gamma",
    link = "log"
)
\end{verbatim}

Bambi specification for binomial model on feedback value for psychological safety:
\begin{verbatim}
feedback_info_psychological = bmb.Model(
    "p(fb_sum, fb_count) ~ "
    "(PERCEIVED_SAFETY + SENSE_OF_CONTROL) * condition",
    df,
    family="binomial"
)
\end{verbatim}

Bambi specification for Bernoulli model on feedback value for physical safety:
\begin{verbatim}
feedback_info_physical = bmb.Model(
    "fb_val ~ (log_max_threat_pr + max_threat_ar) * condition + (1|pid)",
    df,
    family="bernoulli",
)
\end{verbatim}

\begin{table}[pos=htb]
    \caption{Feedback on physical safety, parameter estimates of the Bernoulli model}
    \label{tab:bayes_results_ittc}
    \begin{tabular}{lccccc}
        \toprule
        \textbf{Parameter} & \textbf{Mean} & \textbf{SD} & \textbf{95\% EtiLB} & \textbf{95\% EtiUB} & \textbf{R-hat} \\
        \midrule
        condition[UIF] & -0.21 & 0.81 & -1.8 & 1.3 & 1.00 \\
        max\_threat\_human & -0.114 & 0.083 & -0.29 & 0.039 & 1.00 \\
        max\_threat\_robots & -1.03 & 1.41 & -3.7 & 1.7 & 1.00 \\
        max\_threat\_human:condition[UIF] & 0.16 & 0.091 & -0.01 & 0.35 & 1.00 \\
        max\_threat\_robots:condition[UIF] & -1.12 & 1.47 & -4 & 1.8 & 1.00 \\
        \bottomrule
    \end{tabular}
    \par\smallskip\footnotesize
    Note: UIF: user-initiated feedback (reference = system-initiated feedback); Eti: equal-tailed interval; LB = lower bound, UB = upper bound.
\end{table}

\begin{table}[pos=htb]
    \caption{Feedback on psychological safety, parameter estimates of the binomial model}
    \label{tab:bayes_results_safety}
    \begin{tabular}{lccccc}
        \toprule
        \textbf{Parameter} & \textbf{Mean} & \textbf{SD} & \textbf{95\% EtiLB} & \textbf{95\% EtiUB} & \textbf{R-hat} \\
        \midrule
        condition[UIF] & -1.191 & 0.372 & -1.9 & -0.47 & 1.00 \\
        PERCEIVED\_SAFETY & 0.066 & 0.252 & -0.41 & 0.57 & 1.00 \\
        SENSE\_OF\_CONTROL & 0.357 & 0.303 & -0.24 & 0.94 & 1.00 \\
        PERCEIVED\_SAFETY:condition[UIF] & 1 & 0.439 & 0.13 & 1.9 & 1.00 \\
        SENSE\_OF\_CONTROL:condition[UIF] & -0.34 & 0.404 & -1.1 & 0.46 & 1.00 \\
        \bottomrule
    \end{tabular}
    \par\smallskip\footnotesize
    Note: UIF: user-initiated feedback (reference = system-initiated feedback); Eti: equal-tailed interval; LB = lower bound, UB = upper bound.
\end{table}

\end{document}